\documentclass[prb,showpacs,floatfix,twocolumn,letterpaper,longtable,tabularx]{revtex4}

\usepackage{graphicx,color,float}  

\newcommand{\mno}{Mn$_3$O$_4$}

\begin{document}

\title{Structural change and phase coexistence upon magnetic ordering in the magnetodielectric spinel Mn$_3$O$_4$}

\author{Moureen C. Kemei}\email{kemei@mrl.ucsb.edu}
\author{Jaye K. Harada}\email{jkharada@mrl.ucsb.edu}
\author{Ram Seshadri}\email{seshadri@mrl.ucsb.edu}
\affiliation{Materials Department and Materials Research Laboratory\\
University of California, Santa Barbara, CA, 93106, USA}
\author{Matthew R. Suchomel}\email{suchomel@aps.anl.gov}
\affiliation{X-Ray Science Division and Material Science Division\\ 
Argonne National Laboratory, Argonne IL, 60439, USA}

\date{\today}  

\begin{abstract}
Cooperative Jahn-Teller ordering is well-known to drive the cubic $Fd\overline{3}m$ to tetragonal
$I$4$_1$/$amd$ structural distortion in \mno\, at 1170\,$^{\circ}$C. Further structural distortion
occurs in \mno\, upon magnetic ordering at 42\,K. Employing high-resolution variable-temperature synchrotron
x-ray diffraction we show that tetragonal $I$4$_1$/$amd$ and orthorhombic $Fddd$ phases 
coexist, with nearly equal fractions, below the N\'eel temperature of \mno. Significant variation of the
orthorhombic $a$ and $b$ lattice constants from the tetragonal $a$ lattice constant is observed.
Structural phase coexistence in \mno\, is attributed to large strains due to the lattice mismatch
between the tetragonal $I$4$_1$/$amd$ and the orthorhombic $Fddd$ phases. Strain tensors determined
from Rietveld refinement show a highly strained matrix of the $I4_1/amd$ phase that accommodates the
nucleated orthorhombic $Fddd$ phase in the phase coexistence regime. A comparison of the deformation
observed in \mno\, to structural deformations of other magnetic spinels shows that phase coexistence
may be a common theme when structural distortions occur below 50\,K. 

\end{abstract}

\pacs{75.47.Lx, 75.50.Gg, 61.50.Ks}
\maketitle

\newpage

\section{Introduction}

Phase coexistence is a recurring theme that has been extensively investigated in perovskite
manganese oxides displaying colossal magnetoresistance.\cite{jin_1994} Phase separation of 
charge ordered insulating (antiferromagnetic) and metallic (ferromagnetic) phases has been observed 
in the manganites (La,Pr,Ca)MnO$_3$.\cite{fath_1999,uehara_1999} Competing double exchange mechanisms 
and Jahn-Teller-like electron-lattice coupling have been proposed to explain the coexistence of 
multiple phases.\cite{fath_1999,uehara_1999} Chemical disorder has also been invoked to explain
microscale structural phase separation.\cite{ahn_2004,llobet_2000} However, generalizations in
describing the mechanisms of phase coexistence are often insufficient in describing all aspects of
these phenomena prompting continued interest in these materials.\cite{hwang_1995} Here, we encounter
structural phase coexistence below the N\'eel temperature of the spinel \mno.

Mn$_3$O$_4$ (the mineral hausmanite) consists of edge sharing MnO$_6$ octahedra that are corner 
connected to MnO$_4$ tetrahedra. It is a cubic spinel in the space group $Fd\overline{3}m$ above
1170\,$^{\circ}$C. \mno\, undergoes a cooperative Jahn-Teller distortion below 1170\,$^{\circ}$C due
to orbital degeneracy of the $e_g^1$ states of Mn$^{3+}$ cations that occupy the octahedral sites.
Jahn-Teller ordering of the octahedral $e_g$ states stabilizes the $z^2$ orbital by elongating the
MnO$_6$ octahedral units resulting in tetragonal $I4_1$/$amd$ symmetry below
1170\,$^{\circ}$C.\cite{goodenough_1955, hook_1958} \mno\, is a paramagnet above the
N\'eel temperature ($T_N$)\,=\,41\,K, below $T_N$, it adopts a non-collinear magnetic structure that
consists of ferromagnetically coupled Mn$^{2+}$ cations along the (010) direction and two Mn$^{3+}$
sub-lattices with a net moment that couples antiferromagnetically to the magnetization of Mn$^{2+}$
cations. A transition to an incommensurate spiral configuration occurs at 39\,K, followed by a
transition to commensurate spiral order below 33\,K.\cite{jensen_1974} Commensurate spiral spin
ordering in \mno\, is described by an orthorhombic magnetic unit cell, which is twice the size of
the chemical tetragonal unit cell.\cite{jensen_1974}  Magnon excitations in
ferrimagnetic \mno\, have been recently investigated by Gleason $et\,al.$\cite{gleason_2014} Magnetism in \mno\, couples to its dielectric
properties. Consequently, dielectric anomalies are observed near the magnetic transition
temperatures. In addition, magnetic field control of the dielectric constant below $T_N$ has also
been demonstrated.\cite{tackett_2007,suzuki_2008} 

While Jahn-Teller ordering yields a high temperature cubic-tetragonal distortion in
\mno\,,\cite{goodenough_1955, hook_1958} spin ordering drives further lattice distortions in this spinel.\cite{kim_2010} Kim $et\,al.$ reported a transition from tetragonal to monoclinic
symmetry in single crystal \mno\, at the commensurate spiral ordering temperature ($T$\,$\sim$\,33\,K)
under zero applied fields.\cite{kim_2010}  In a subsequent report, Kim and coauthors proposed that
the low temperature structure of \mno\, is orthorhombic in the space group $Fddd$.\cite{kim_2011}
Similarly, Chung $et\,al.$ have recently reported orthorhombic instabilities in
\mno.\cite{chung_2013} While these initial efforts clearly illustrate a structural transformation in
ferrimagnetic \mno, the complete low-temperature structure of \mno\, remains unresolved and is the
focus of the present work. Extensive studies of the magnetostructural phases of \mno\, by Kim and
coauthors show that in the presence of large magnetic fields, the structural distortion occurs at
higher temperatures in the incommensurate spiral magnetic phase. Remarkably, a spin disordered phase
can be stabilized in \mno\, near 0\,K when intermediate fields are applied parallel to the
(1$\overline{1}$0) direction. Here, applied fields transverse to the magnetic ordering direction
frustrate spin ordering resulting in disordered spins far below the magnetic ordering
temperature.\cite{kim_2010}  Kim $et\,al.$ have also shown an increase in the N\'eel temperature
under pressure to temperatures as high as 54\,K.\cite{kim_2011} 

We present here, a complete description of the low-temperature nuclear structure of \mno\, 
finding that tetragonal $I$4$_1/amd$ and orthorhombic $Fddd$ phases coexist in ferrimagnetic \mno. 
The orthorhombic $Fddd$ phase is spontaneously nucleated at the N\'eel temperature and its phase 
fraction increases slightly, attaining a maximum of about 55$\%$ near 8\,K. The evolution of the 
unit cell volume as a function of temperature indicates a distortion in both the tetragonal 
$I$4$_1$/$amd$ and orthorhombic $Fddd$ phases below 42\,K. Distortions in both low temperature 
tetragonal and orthorhombic phases are corroborated by detailed studies of polyhedra distortions 
which show deformations in both of these phases. In the orthorhombic phase, MnO$_6$ octahedra show elongation of
equatorial Mn-O bonds while the MnO$_4$ tetrahedra show a decrease in the Mn-O bond length.
Conversely, the distorted tetragonal phase shows shortened Mn-O equatorial bonds in the MnO$_6$
octahedra and elongated Mn-O bonds in the MnO$_4$ tetrahedra. We examine the role of strain in
stabilizing coexisting tetragonal and orthorhombic phases in ferrimagnetic \mno\, and make
comparisons of the structural distortion of \mno\, to those of other magnetic spinels. The
complexities in understanding the variations of structural deformations in magnetic spinels are
highlighted.  The complete structural description of this spinel is pivotal to unraveling the
complex ground states of this material. These results necessitate a reinvestigation of the
magnetic structure of \mno,  which was resolved considering only the tetragonal nuclear
structure below the N\'eel temperature. 

\section{Methods}

Polycrystalline \mno\, was prepared from a MnC$_2$O$_4\cdot$2H$_2$O precursor. The oxalate was decomposed
at 600\,$^{\circ}$C for 3 hrs. The precursor powder was then ground, pelletized, and annealed at
1000\,$^{\circ}$C for 24\, hrs and water quenched. Variable-temperature (7.5\,K -- 450\,K) 
high-resolution ($\delta Q$/$Q\,\leq\,$2$\,\times$10$^{-4}$, $\lambda$\,=\,0.41394\,\AA) synchrotron
x-ray powder diffraction was performed at beamline 11BM of the Advanced Photon Source, Argonne National
Laboratory and the ID31 beamline ($\lambda$\,=\,0.39985\,\AA) of the European Synchrotron
Radiation Facility. Variable temperature measurements were collected on heating with a temperature scan
rate of 0.8\,K$/$min and an x-ray pattern was collected every 3 minutes. Diffraction patterns were fit to
crystallographic models using the EXPGUI/GSAS software program.\cite{toby_expgui_2001,larson_2000}
Crystal structures were visualized using the program VESTA.\cite{momma_vesta_2008} Density measurements of a powder sample of \mno\, were performed on a MicroMetrics AccuPyc II 1340 pycnometer. A sample cup with a volume of 0.1\,cm$^3$ was filled with \mno\, powder with a mass of 93.1\,mg during the density measurement. Temperature-dependent
and field-dependent magnetic measurements were performed using the Quantum Design (QD) MPMS 5XL
superconducting quantum interference device (SQUID). Heat capacity measurements were collected using a QD
Physical Properties measurement system.  \mno\, pellets for dielectric measurements were spark plasma
sintered at 1000\,$^{\circ}$C under a 6\,kN load for 3 minutes. Prior to measuring dielectric properties,
the spark plasma sintered pellet was annealed at 1000\,$^{\circ}$C for 12\,hrs, quenched, and
characterized by x-ray diffraction to ensure that stoichiometric \mno\, was retained. Dielectric
measurements were performed on a pellet with a diameter of 9.51\,mm and a thickness of 2.28\,mm whose
cylindrical faces were coated by conducting epoxy. Dielectric properties were measured by an
Andeen-Hagerling AH2700A capacitance bridge connected to the sample by shielded coaxial cables. The
sample was placed in a QD Dynacool Physical Properties measurement system which provides a variable
temperature and variable field environment when carrying out dielectric measurements.

\section{Results and Discussion}

\begin{figure} [h!]
\centering \includegraphics[width=3.4in]{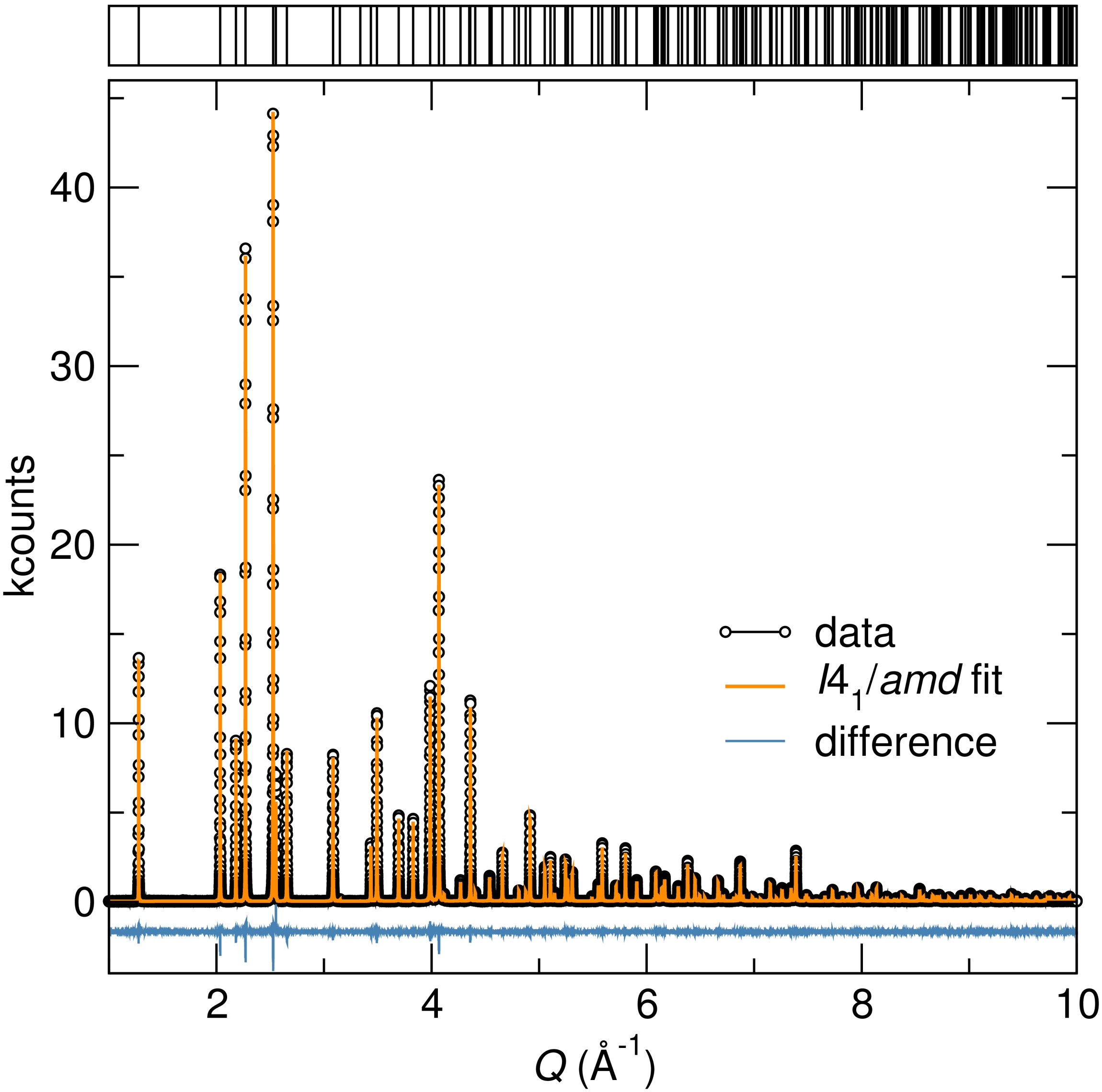}\\
\caption{(Color online) The room temperature synchrotron diffraction pattern of \mno\, modeled using
the tetragonal $I$4$_1$/$amd$ structure. 
\label{fig:RT}}
\end{figure}

At room temperature \mno, is a tetragonal spinel in the space group $I$4$_1$/$amd$. The synchrotron x-ray
diffraction pattern measured at 298\,K and modeled to the space group $I$4$_1$/$amd$ is illustrated in
fig. \ref{fig:RT}. Rietveld refinement of the diffraction pattern yields lattice constants
$a$\,=\,5.76289(2)\,\AA\, and $c$\,=\,9.46885(1) \,\AA\, with a $c$/$a\,\sqrt{2}$ of 1.16 in good
agreement with values from the literature.\cite{shoemaker_2009,chardon_1986} Goodness of fit parameters $\chi^2$,
R$_{wp}$, and R$_p$ of 2.354, 10.23\,$\%$, and 8.11\,$\%$ respectively are obtained from the refinement.
Valence bond sums computed using room temperature bond lengths show that \mno\, is a normal spinel with
tetrahedral and octahedral valence states of 2.01 and 3.02 respectively. 

\begin{figure} 
\centering \includegraphics[width=3.4in]{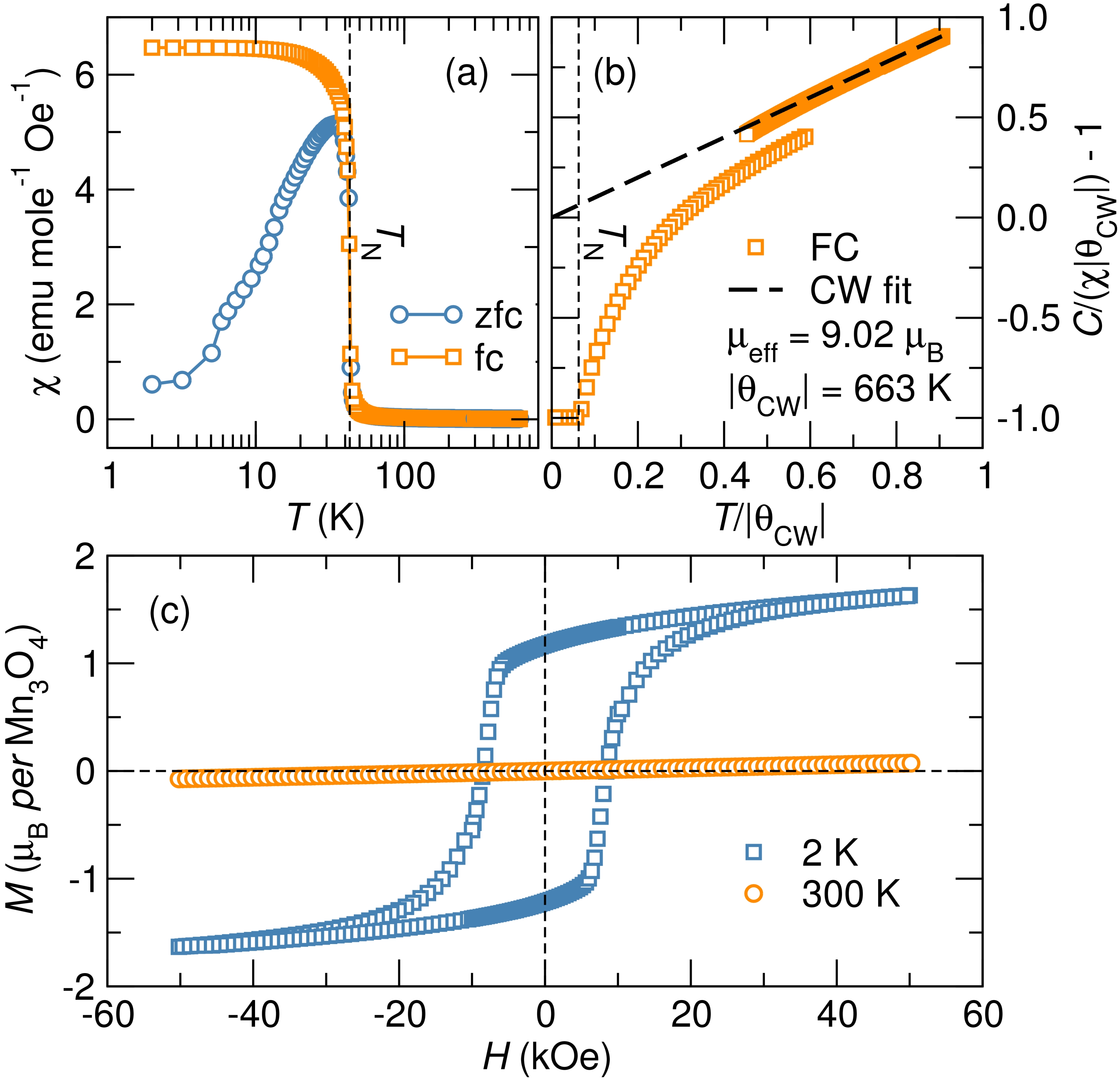}\\ 
\caption{(Color online) (a)
Temperature dependent magnetic susceptibility measurements of \mno\, performed in a 1000\,Oe field show a
deviation of the zero-field cooled susceptibility from the field cooled susceptibility below the N\'eel
temperature $T_N$\,=\,42\,K. (b) Curie-Weiss fitting in the temperature range
500\,K\,$<$\,$T$\,$<$\,600\,K yields an effective moment of 9.02\,$\mu_B$ and an expected ordering
temperature $|\Theta_{CW}|$\,=\,663\,K. High-temperature and low-temperature susceptibility measurements
were conducted on different instruments and instrumental variations yield the slight offset between the
low-temperature and high-temperature susceptibility that is negligible in plot (a) but emphasized in the
$1/\chi$ scaling of plot (b). (c) Field-dependent magnetization shows paramagnetic behavior above the
N\'eel temperature of \mno\, and ferrimagnetic behavior below $T_N$ with a saturation magnetization of
1.63\,$\mu_B$ and a coercive field of 8.5\,kOe when $H$\,=\, 50\,kOe and $T$\,=\,2\,K.    
\label{fig:Mag}} \end{figure}

A broad transition from a paramagnetic to a ferrimagnetic state, which is characteristic of
polycrystalline \mno\,,\cite{tackett_2007} occurs near $T_N$\,=\,42\,K. At $T_N$, a separation between
the zero-field cooled and field cooled curves develops and is enhanced with decrease in temperature [fig.
\ref{fig:Mag} (a)]. While a single broad magnetic transition is observed in the temperature-dependent
susceptibility, the magnetic structure of \mno\, is extremely rich, featuring a transition to a collinear
magnetic structure at 42\,K followed by the onset of an incommensurate magnetic spiral at 39\,K, and
finally a commensurate spiral magnetic state occurs below 33\,K.\cite{jensen_1974,boucher_1971} Each of
these transitions is observed in heat capacity and capacitance measurements as discussed later in the
manuscript. Curie-Weiss fitting in the temperature range 500\,K\,$<$\,$T$\,$<$\,600\,K yields a
$\Theta_{CW}$ of 663\,K in good agreement with prior work [fig. \ref{fig:Mag} (b)].\cite{chardon_1986,
boucher_1971} Comparison between the expected Curie-Weiss ordering temperature and $T_N$ results in a
significant frustration index of 15.8. This illustrates that frustration may play a role in the magnetism
of \mno. The Curie-Weiss fit also yields an effective moment of 9.02\,$\mu_B$ which is congruent with the
expected effective moment of 9.11\,$\mu_B$ computed from the spin-only effective moments of Mn$^{2+}$ and
Mn$^{3+}$ which are 5.92\,$\mu_B$ and 4.9\,$\mu_B$ respectively. Short-range correlations cause
deviations of the inverse susceptibility from the Curie-Weiss model above the N\'eel temperature as
illustrated in fig.\ref{fig:Mag} (b). 

\begin{figure}
\centering \includegraphics[width=3.4in]{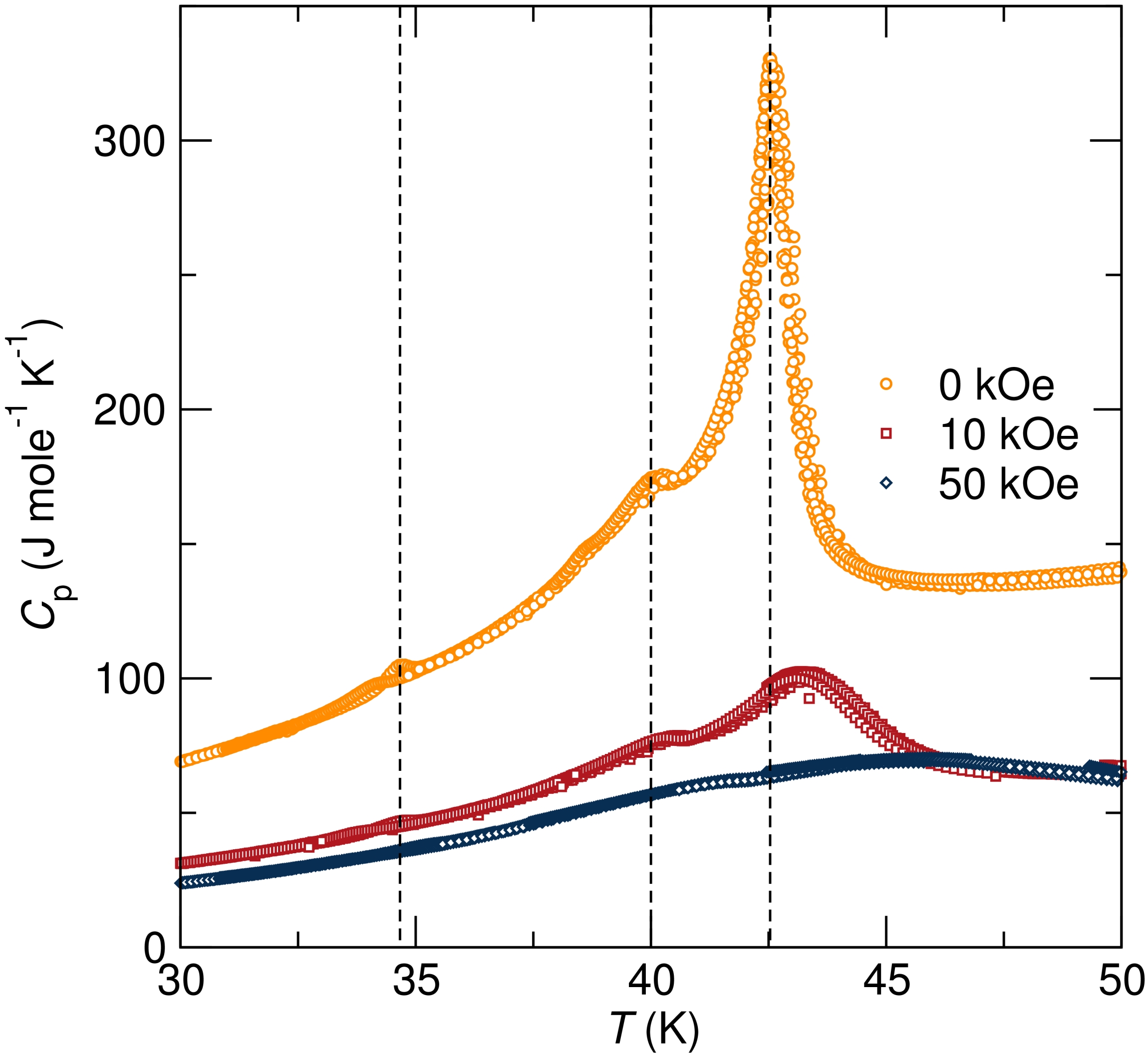}\\
\caption{(Color online) Heat capacity measurements of \mno\, show three anomalies associated with changes
in magnetic structure at 42.5\,K, 40\,K, and 34.5\,K. These transitions are evident under zero field
conditions but they are suppressed and shifted to higher temperatures in the presence of a magnetic
field. 
\label{fig:HC}}
\end{figure}

The weak linear increase in magnetization with field in the paramagnetic phase evolves to a hysteretic
magnetization in ferrimagnetic \mno\,[fig. \ref{fig:Mag} (c)]. A saturation magnetization of
1.63\,$\mu_B$ and a coercive field of 8.5\,kOe are measured at 50\,kOe and 2\,K. A collinear
ferrimagnetic state in \mno\, would yield a saturation magnetization of 3\,$\mu _B$/formula unit,
however, the measured saturation moment of 1.63\,$\mu _B$ is consistent with the reported spiral magnetic
structure of \mno\, near 2\,K.\cite{jensen_1974} \mno\, exhibits significant anisotropy with an easy axis
along the (001) direction, which yields a spontaneous magnetization of 1.85\,$\mu_B$ in single
crystalline samples.\cite{dwight_1960} However, a slightly decreased value of the spontaneous
magnetization is expected in polycrystalline materials due to the random alignment of \mno\,
grains.\cite{dwight_1960, tackett_2007} 

Spin ordering in \mno\, leads to changes in entropy that are illustrated in fig. \ref{fig:HC}. Variations
in magnetic structure give rise to distinct heat capacity anomalies at 42.5\,K, 40\,K, and 34.5\,K under
zero-field conditions. The largest change in entropy occurs at 42\,K where the highest heat capacity peak
is observed. Entropy changes in \mno\, depend on the applied magnetic field, with large fields
suppressing the heat capacity transitions. At 10\,kOe field, heat capacity peaks are broad but still
visible at the transition temperatures and the 42.5\,K peak shifts to a higher temperature, $T_N$\,=\,
43.2\,K. Pronounced suppression of the heat capacity transitions is evident at 50\,kOe. The trend in the
field-dependent heat capacity reported here, namely, the suppression of heat capacity anomalies and the
shift to higher temperatures is in agreement with the work of Kim $et\,al.$, which reports an increase in
$T_N$ in the presence of a field and complete suppression of the onset of magnetic order when a magnetic
field is applied along certain crystallographic directions.\cite{kim_2010, kim_2011} 

\begin{figure}
\centering \includegraphics[width=3.4in]{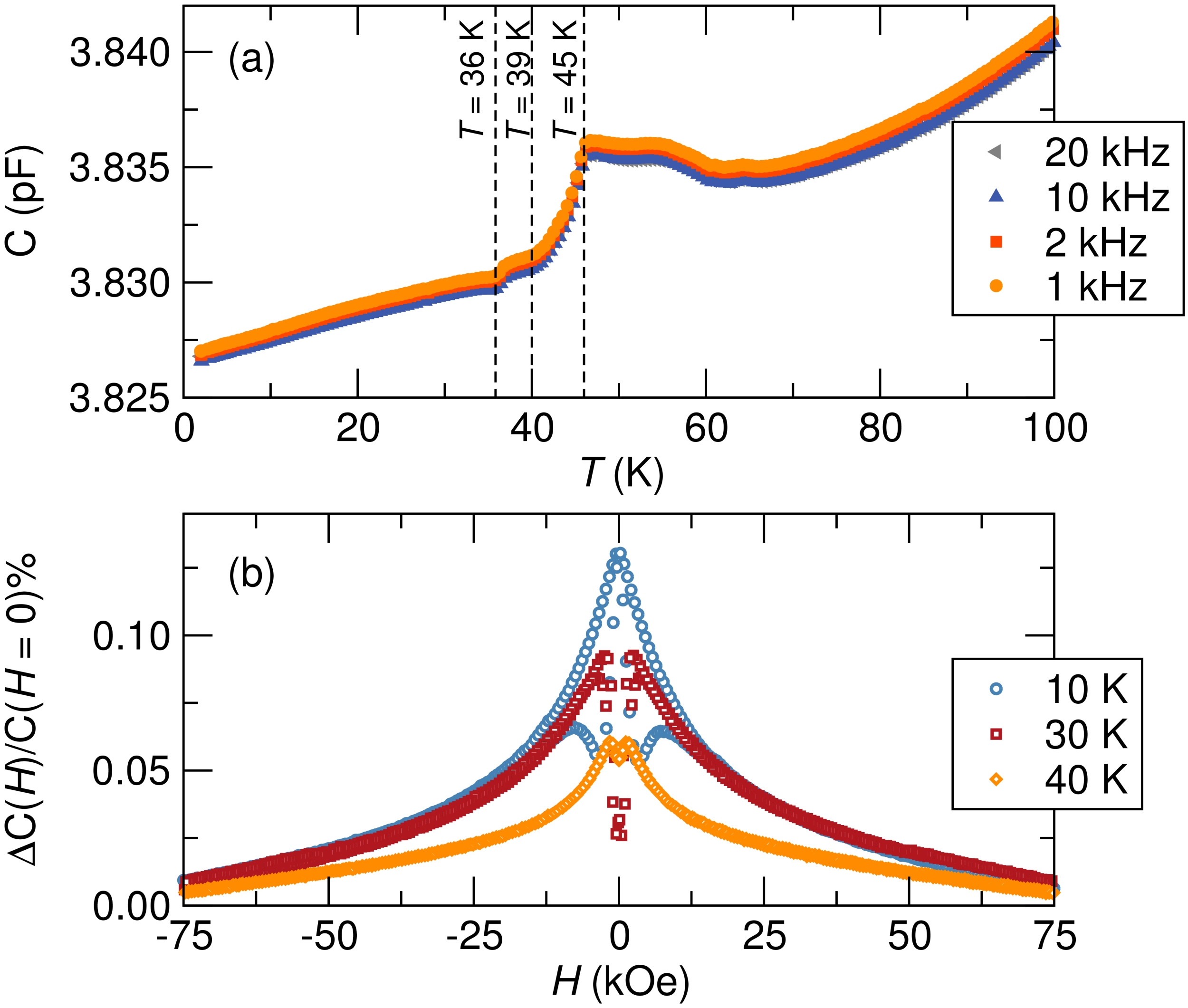}\\
\caption{(Color online) Magnetodielectric coupling in \mno. (a) The dielectric constant as a function of
temperature shows three anomalies at 36\,K, 39\,K, and 45\,K which are in close proximity to spin
ordering transitions. (b) A magnetic field can modulate the dielectric constant below $T_N$. The changes
in the dielectric constant in the presence of a magnetic field increase with decrease in temperature. 
\label{fig:dielectric constant}}
\end{figure}

\mno\, is a known magnetodielectric.\cite{tackett_2007,suzuki_2008} Figure \ref{fig:dielectric constant}
(a) shows anomalies in the dielectric constant occurring in close proximity to spin ordering transition
temperatures at 45\,K, 39\,K, and 36\,K. The dielectric constant shows no frequency dependence and this illustrates 
the intrinsic nature of magnetodielectric coupling in the studied \mno\, sample. A frequency
dependent dielectric constant would indicate that magnetoresistive effects are contributing to
magnetodielectric coupling. We also note the presence of dielectric anomalies above 45\,K[fig \ref{fig:dielectric constant} (a)], which could be linked to short-range spin correlations in the temperature range $T_N<$\,$T$\,$<$\,$|\Theta_{CW}|$. The origin of the dielectric changes above $T_N$ should be further investigated. The dielectric constant exhibits strong field dependence below $T_N$ [fig.\ref{fig:dielectric constant} (b)]. The field dependence of the dielectric constant increases with
decrease in temperature.  

\begin{figure} \centering \includegraphics[width=3.4in]{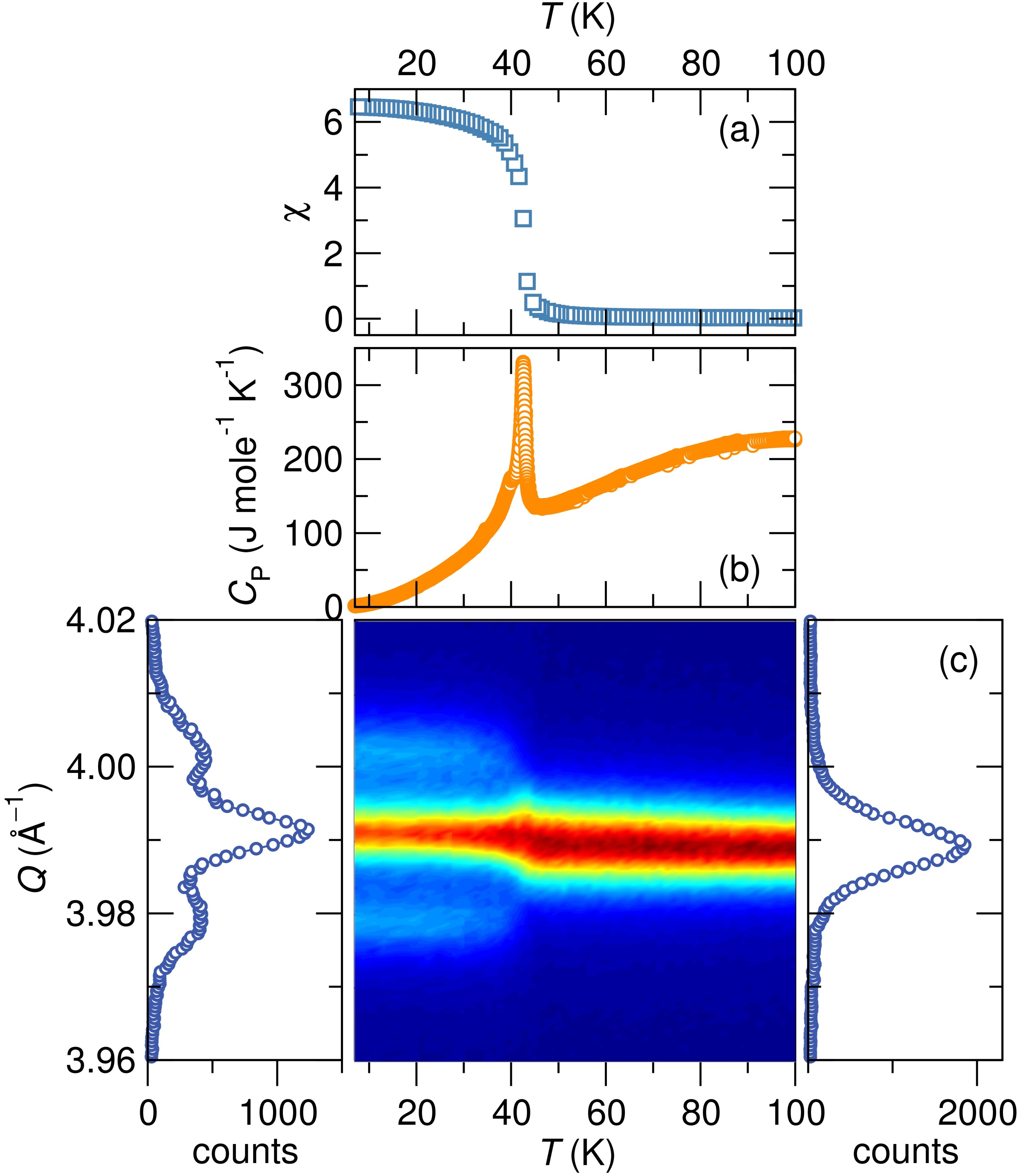}\\ \caption{(Color online)
Concurrent magnetic and structural ordering in \mno. (a) Magnetic susceptibility measurements in units of
emu mole$^{-1}$ Oe$^{-1}$ and (b) heat capacity measurements show the onset of long-range ferrimagnetic
order in \mno\, near 42\,K. (c) A structural distortion occurs concomitantly with the onset of
ferrimagnetic order in \mno\, at 42\,K where variable-temperature high-resolution synchrotron x-ray
diffraction shows a splitting of the high temperature tetragonal (321) reflection. While there is a
decrease in intensity of the tetragonal (321) reflection in the low temperature phase, we note that it
coexists with the emergent peaks to the lowest temperatures studied. \label{fig:VT}} \end{figure} 

The onset of long-range magnetic order in \mno\, occurs concurrently with a structural distortion. Figure
\ref{fig:VT} (a) and (b) show spin and entropy changes occurring concurrently near 42\,K. Figure
\ref{fig:VT} (c) also shows x-ray diffraction evidence of structural changes in \mno\, occurring simultaneously near 42\,K. A
splitting of the tetragonal $I4_1/amd$ (321) reflection to several peaks below 42\,K is clearly
demonstrated by variable-temperature high-resolution synchrotron x-ray diffraction [Fig. \ref{fig:VT}
(c)]. The emergence of new diffraction peaks below 42\,K indicates an average structure distortion. This
study of polycrystalline \mno\, shows that the magnetostructural distortion occurs near 42\,K [fig.
\ref{fig:VT} (c)] while earlier studies of single crystal \mno\, have reported a 33\,K magnetostructural
distortion under zero-field conditions and a 39\,K spin-drive structural transition in the presence of an applied magnetic
field.\cite{kim_2010, kim_2011} The difference in the transition temperature is attributed to varying
strain effects in polycrystalline versus single crystal samples; large strains in single crystal samples
could suppress the onset of the structural transition. In addition, the cooling and heating rates of the
sample during a structural study are also expected to influence the structural distortion temperature. 

\begin{figure} 
\centering \includegraphics[width=3.4in]{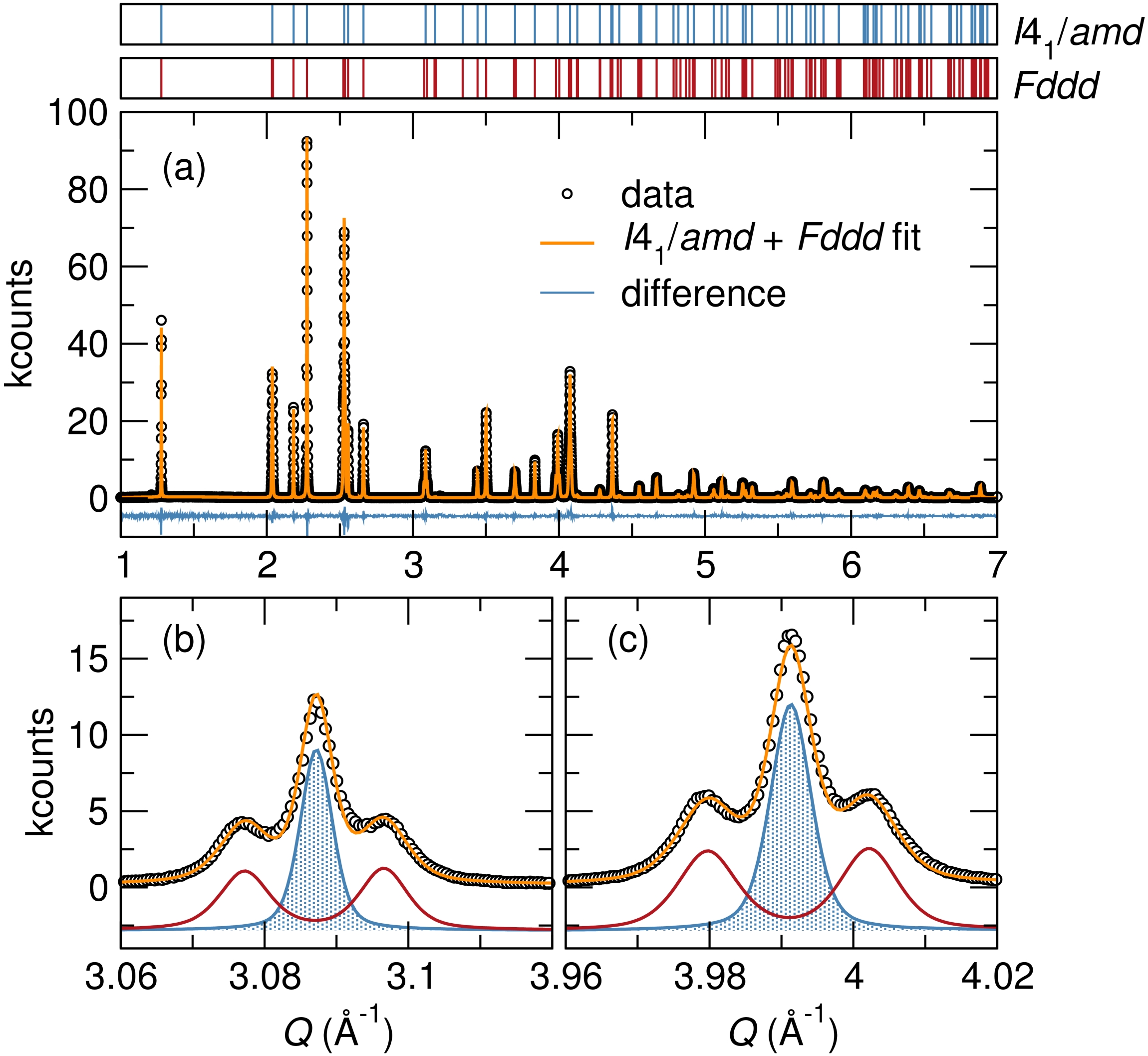}\\ 
\caption{(Color online) (a) The
8.2\,K synchrotron x-ray diffraction pattern of \mno\, modeled to a structure of coexisting tetragonal
$I4_1/amd$ and orthorhombic $Fddd$ structures. (b) The tetragonal $I4_1/amd$ (220) reflection splits into
three reflections consisting of the tetragonal reflection and the orthorhombic $Fddd$ reflections (040)
and (400). (c) In a similar way, tetragonal (321) reflection splits giving rise to the orthorhombic
$Fddd$ (511) and (151) reflections.  \label{fig:LT}} 
\end{figure} 

\begin{table}
\caption{\label{tab:rietveldmno} Structural parameters of 
coexisting tetragonal $\textit{I}$4$_1$/$\textit{amd}$ and orthorhombic $\textit{Fddd}$ phases of \mno\, at 8.2\,K.}
\centering
\begin{tabular}{ll}
\hline
\hline
Temperature (K) & 8.2 \ \\
Space group	&$\textit{I}$4$_1$/$\textit{amd}$\ \\
Setting 		& origin 2 \ \\
$Z$  		& 4 \ \\
$a$(\AA) 		&5.75638(1)\ \\
$c$(\AA) 		&9.44393(4)\ \\
Vol/(\AA$^3$)& 312.934(1)\ \\
Mn$^{2+}$	& $4a$ (0,\,1/4, \,7/8)\ \\
Mn$^{3+}$ 	& $8d$ (0,\,1/2, \,1/2) \ \\
O 			& $16h$ ($0,y,z$)\ \\
 			& $y$ 0.4715(2)\ \\
 		        & $z$ 0.2595(2)\ \\
Mn$^{2+}$ $U_{iso}$ ($10^{-2}$ \AA$^2$) 	& 1.41(2)\ \\
Mn$^{3+}$  $U_{iso}$ ($10^{-2}$ \AA$^2$) 	& 1.38(1)\ \\
O $U_{iso}$ ($10^{-2}$ \AA$^2$) 	& 1.51(3)\ \\
Wt. frac.		&	0.44(6)\ \\
Space group	& $\textit{Fddd}$\ \\
Setting 	       	& origin 2 \ \\
$Z$  		& 8 \ \\
$a$(\AA) 	        &8.11602(5)\ \\
$b$(\AA) 	        &8.16717(5)\ \\
$c$(\AA) 		&9.44209(5)\ \\
Vol/(\AA$^3$)			& 625.869(4)\ \\
Mn$^{2+}$				& $8a$ (1/8,\,1/8, \,1/8)\ \\
Mn$^{3+}$ & $16d$ (1/2,\,1/2, \,1/2) \ \\
O 						& $32h$ ($x,y,z$)\ \\
						& $x$ 0.4871(2)\ \\
 						& $y$ 0.4869(2)\ \\
 		                & $z$ 0.2585(2)\ \\
Mn$^{2+}$ $U_{iso}$ ($10^{-2}$ \AA$^2$) 	& 1.25(2)\ \\
Mn$^{3+}$  $U_{iso}$ ($10^{-2}$ \AA$^2$) 	& 1.21(2)\ \\
O $U_{iso}$ ($10^{-2}$ \AA$^2$) 	& 1.45(3)\ \\
Wt. frac.				&	0.56(6)\ \\
$\chi^2$ 				& 3.206 \ \\
$R_p$($\%$)				& 5.46\ \\
$R_{wp}$($\%$)			& 6.82\ \\
\hline
\hline
\end{tabular}
\end{table}

The precise structural description of \mno\, in the ferrimagnetic state is so far unknown.  We find that
\mno\, undergoes a phase transformation from tetragonal $I4_1/amd$ symmetry to a phase coexistence regime
consisting of both tetragonal $I4_1/amd$ and orthorhombic $Fddd$ phases. Figure \ref{fig:LT} (a) shows
the 8.2\,K diffraction pattern of \mno\, modeled by tetragonal $I4_1/amd$ and orthorhombic $Fddd$ phases.
Structural parameters and goodness of fit parameters obtained from Rietveld refinement are tabulated in
Table \ref{tab:rietveldmno}. Small goodness of fit parameters along with the agreement between the model
and the data illustrates that coexisting tetragonal and orthorhombic phases characterize ferrimagnetic
\mno. No single phase low symmetry solution could be obtained to model the low temperature structure of \mno. A closer look at some of the diffraction reflections that split below the N\'eel temperature is
presented in  Fig. \ref{fig:LT} (b) where the tetragonal (220) reflection coexists with orthorhombic
(040) and (400) reflections. Similarly, fig. \ref{fig:LT} (c) shows the coexistence of the tetragonal
(321) and the orthorhombic (511) and (151) reflections. The orthorhombic $Fddd$ space group is a subgroup
of the tetragonal space group $I4_1/amd$, and it has been shown to describe structural ground states of
the Jahn-Teller active spinels NiCr$_2$O$_4$ and CuCr$_2$O$_4$.\cite{suchomel_2012} Magnetostructural
phase transitions leading to complex structural ground states are an emerging theme in magnetic spinels;
a recent report from our group shows structural phase coexistence in MgCr$_2$O$_4$ and
ZnCr$_2$O$_4$.\cite{kemei_2013}

\begin{figure} 
\centering \includegraphics[width=3.2in]{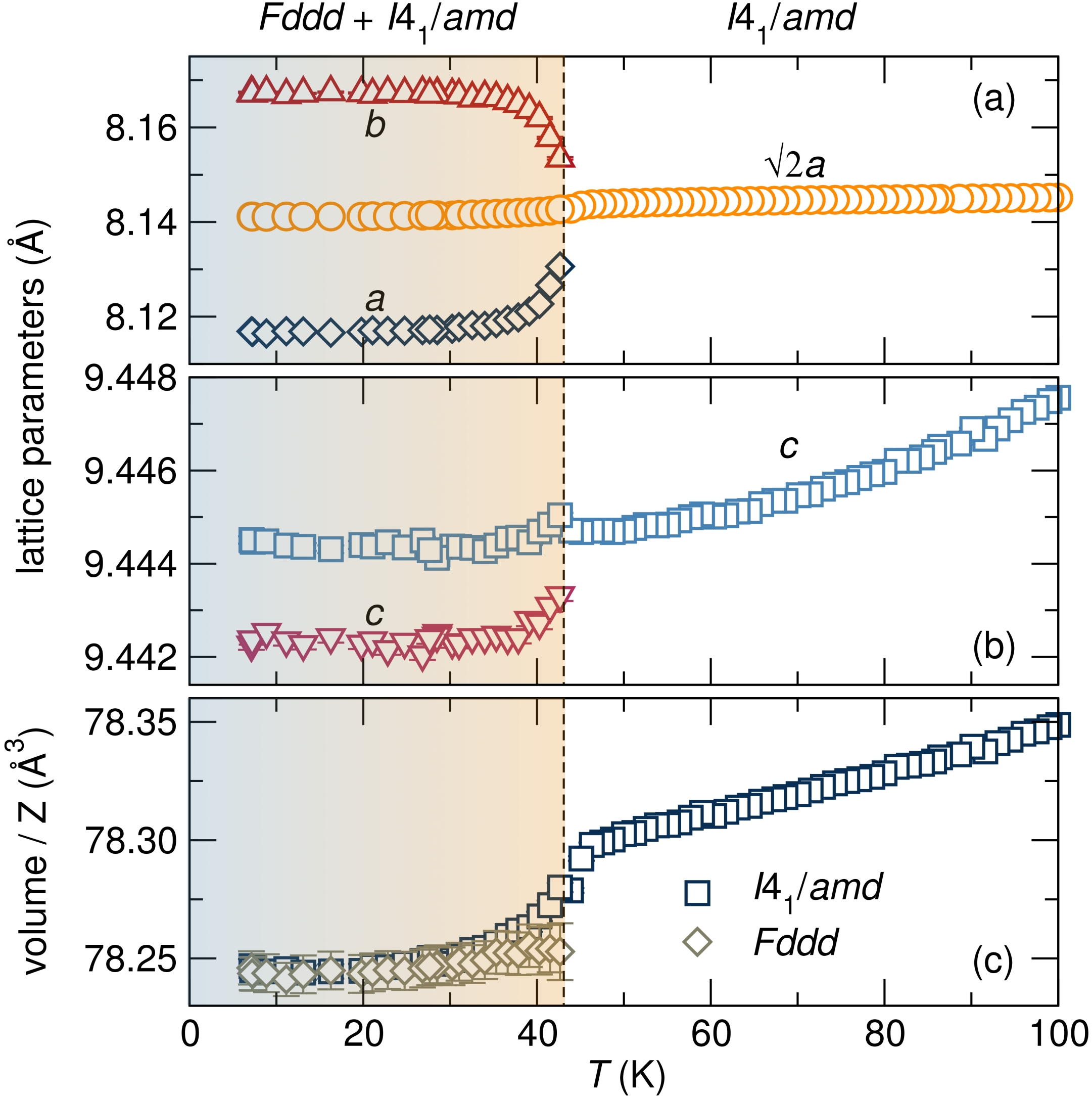}\\ 
\caption{(Color online)
Temperature-dependence of the structural parameters of \mno. (a) Variation of the tetragonal $a$ lattice
constant of \mno\, with temperature shows the emergence of a compressed orthorhombic $a$ lattice constant
and an elongated orthorhombic $b$ lattice constants below 42\,K. (b) The tetragonal $c$ lattice constant
decreases with temperature to 42\,K, below which the orthorhombic $c$ lattice constant, which has a
decreased value, emerges. (c) A linear decrease in the unit cell volume is observed above 42\,K. At
42\,K, a discontinuous decrease in cell volume is observed in both the tetragonal $I4_1/amd$ and
orthorhombic $Fddd$ phases.  \label{fig:LatParam}} 
\end{figure} 
 
The low-temperature structures coexist with nearly equal fractions below $T_N$.  Weight fractions of
50.4\,$\%$ and 49.6\,$\%$ for the $I4_1/amd$ and $Fddd$ phases respectively are measured near $T_N$.
With decrease in temperature below $T_N$, the orthorhombic phase fraction increases slightly attaining a
maximum value of 56\,$\%$ near 8\,K. Scherrer analysis reveals that large crystallite sizes with
dimensions $>$\,10$\mu m$ are observed in the high temperature and low temperature tetragonal phases
while smaller domain sizes of about 690\,nm characterize the orthorhombic phase.  The simple Scherrer
model of crystallite size analysis neglects instrumental broadening, therefore, the sizes obtained are
the minimum values. 

At the magnetostructural ordering temperature of \mno, orthorhombic $Fddd$ lattice constants emerge and
coexist with tetragonal $I4_1/amd$ lattice constants (Fig. \ref{fig:LatParam}). The orthorhombic $a$ and
$b$ lattice constants derive from the tetragonal $a$ lattice parameter. A $\approx$\,0.3\,$\%$
compression of the orthorhombic $a$ axis and a  $\approx$\,0.33\,$\%$ elongation of the orthorhombic $b$
axis relative to the tetragonal $a$ lattice constant are measured [Fig. \ref{fig:LatParam} (a)]. The
continuous decrease of the tetragonal $c$ axis that occurs above 42\,K, is disrupted at $T_N$ where the
orthorhombic $c$ axis emerges with values $\approx$\,0.02\,$\%$ smaller relative to the tetragonal $c$
axis [Fig. \ref{fig:LatParam} (b)]. The temperature-dependence of the unit cell volume shows a
discontinuous decrease below 42\,K [Fig. \ref{fig:LatParam} (c)]. In the Ehrenfest classification of
phase transitions, first order phase transitions are characterized by a discontinuous change in the first
derivative of the free energy. The discontinuous change in the cell volume of \mno\, and coexistence of
two low-temperature structures suggests that this is a first order phase transition. The structural
distortion of \mno, $b_{orth}/a_{tet}$\,=1.003, is of the same order as those of other spin driven
lattice distortions observed in the related spinel compounds NiCr$_2$O$_4$,\cite{suchomel_2012}
CuCr$_2$O$_4$,\cite{suchomel_2012} ZnCr$_2$O$_4$,\cite{kemei_2013} and MgCr$_2$O$_4$.\cite{kemei_2013}

\begin{figure}
\centering \includegraphics[width=3.4in]{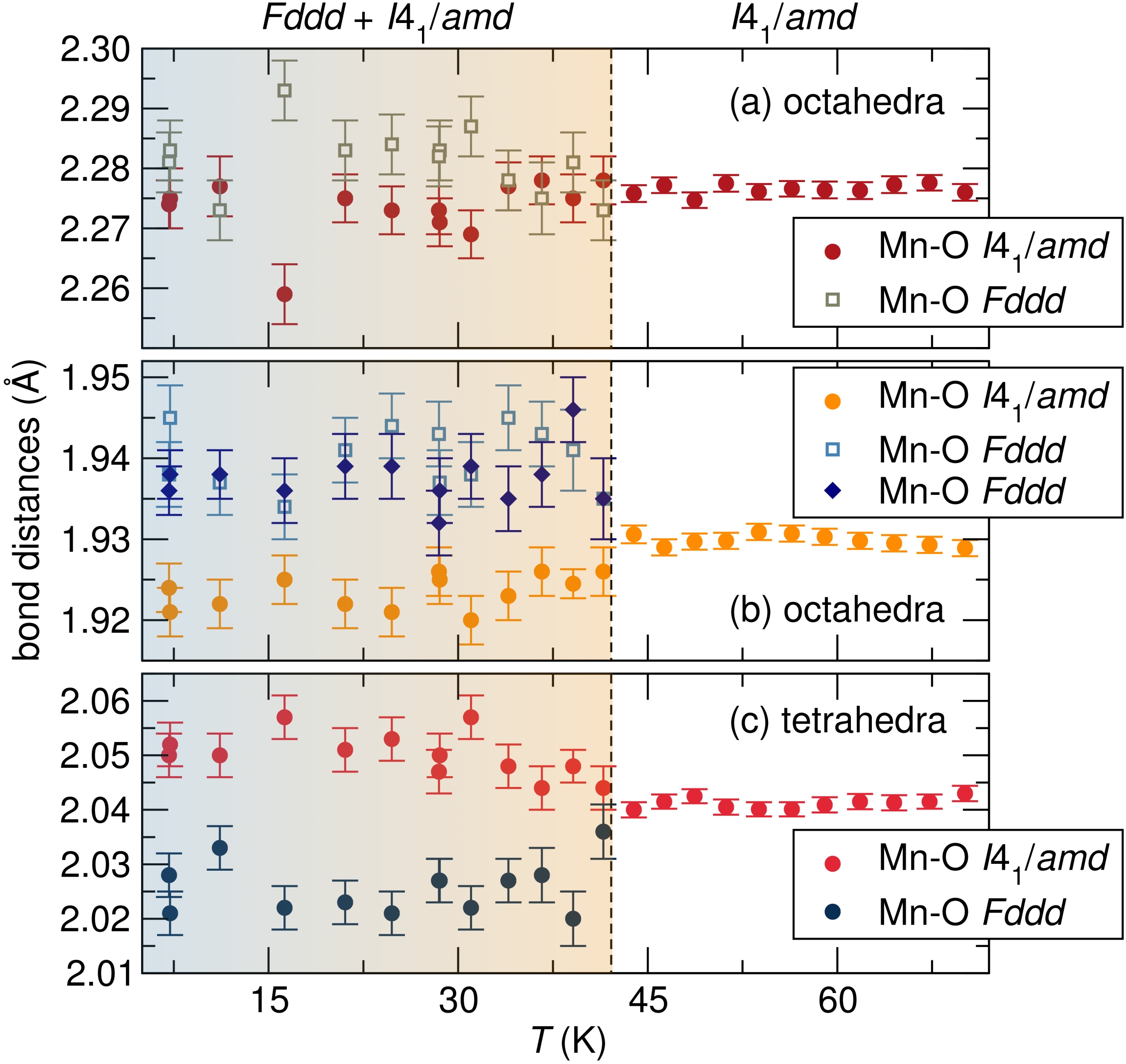}\\
\caption{(Color online) Polyhedra distortions of \mno. (a) The axial Mn-O bond length of the MnO$_6$
octahedra shows relative temperature independence across the structural phase transition in both the
tetragonal and orthorhombic phases. (b) The equatorial Mn-O bond length of MnO$_6$ octahedra in the high
temperature tetragonal phase has a value of about 1.93\,\AA. Below the structural phase transition, this
equatorial bond length is compressed in the tetragonal phase and has a value of $\approx$\,1.92\,\AA. In
the orthorhombic phase, this bond length is distorted giving rise to two different equatorial bonds with
lengths of about 1.94\,\AA. (c) Below $T_N$, the Mn-O bond length of the MnO$_4$ polyhedra increases in
length in the tetragonal phase while its length decreases in the orthorhombic phase.
\label{fig:polydist}}
\end{figure} 

The complete crystallographic description of \mno\, enables detailed investigation of polyhedral
distortions that occur following the structural distortion. The elongated MnO$_6$ polyhedral units of
tetragonal \mno\, above $T_N$ are described by an apical bond length of 2.275(9)\,\AA, which is twofold
degenerate and an equatorial bond length of 1.93(3)\,\AA\, that is fourfold degenerate [Fig.
\ref{fig:polydist} (a) and (b)]. Below the structural transition temperature, the apical bond length
remains fairly temperature independent with small variations around the 2.275(5)\,\AA\, value in both the
tetragonal and orthorhombic phases[Fig. \ref{fig:polydist} (a)]. On the other hand, below $T_N$, the
equatorial bond length of the tetragonal phase decreases to a length of about 1.92(5)\,\AA\, while
retaining its fourfold degeneracy. The equatorial Mn-O bond length of MnO$_6$ octahedra exhibits
distortions in the orthorhombic phase that yield two Mn-O bond lengths that are about 1.935(5)\,\AA\,
long, each with twofold degeneracy [Fig. \ref{fig:polydist} (b)]. The high temperature tetragonal phase
has a tetrahedral Mn-O bond length of 2.04\,\AA\,, this bond length increases to
$\approx$\,2.05(3)\,\AA\, in the low temperature tetragonal phase while it decreases in the orthorhombic
phase to $\approx$\,2.025(4)\,\AA\,[Fig. \ref{fig:polydist} (c)]. The distortions of the polyhedra in
each of the low temperature structures are complementary, for instance examining the low temperature
orthorhombic phase: the elongation of the equatorial bond lengths of MnO$_6$ octahedra are compensated by
the decrease in Mn-O bond lengths of the MnO$_4$ tetrahedra. 

\begin{figure}
\centering \includegraphics[width=3.4in]{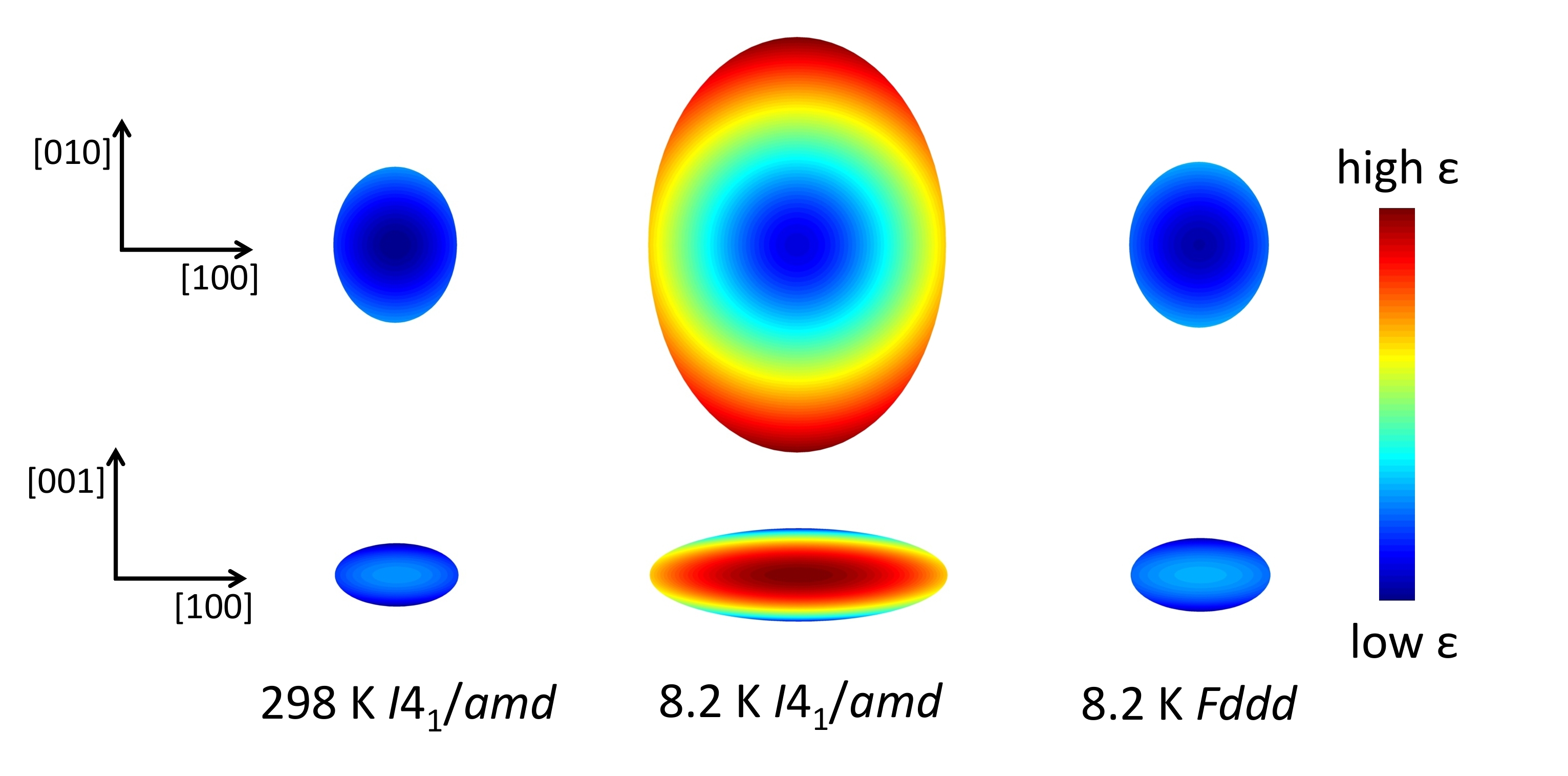}\\
\caption{(Color online)  Ellipsoid representation of the second rank strain tensor of \mno\, in the
various phases as determined by Rietveld refinement using the Pseudo-Voigt with Finger-Cox-Jephcoat
asymmetry(GSAS profile 3) profile function. In addition to the color scale, the size of the ellipsoid
scales with the magnitude of the strain. In all the phases of \mno, large strains are observed in the
[110] plane compared to the [001] direction. The low temperature tetragonal $I4_1/amd$ phase is highly
strained compared to the high temperature tetragonal phase and low temperature orthorhombic phase. The
high temperature  $I4_1/amd$ phase shows the lowest strains of the three phases of \mno. 
\label{fig:strain}}
\end{figure} 

\begin{table}[h!] 
\caption{\label{tab:microstrain} Microstrain terms of the various phases of \mno\, obtained from 
Rietveld refinement. 
The magnitude of the microstrain terms of the low temperature tetragonal phase are larger than those of the other 
phases of \mno. These strain tensors have been plotted in an ellipsoidal representation in 
Fig \ref{fig:strain}. Pseudo-Voigt profile with Finger-Cox-Jephcoat asymmetry was employed in 
describing the diffraction lineshapes.}
\centering
\begin{tabular*}{8.8 cm}{|l|l|l|l|}
\hline
Microstrain term & 298\,K $\textit{I}$4$_1$/$\textit{amd}$ &  
8.2\,K $\textit{I}$4$_1$/$\textit{amd}$ & 8.2\,K $Fddd$ \ \\
\hline
\hline
$L_{11}$ & 0.1071    & $-$0.0257   & 0.1666\\
$L_{22}$ & 0.159     & $-$0.3624   & 0.1975\\
$L_{33}$ & 0.0743    & 0.1071      & 0.0837\\
$L_{12}$ & 0.1093    & 0.3676      & $-$0.0481\\
$L_{13}$ & $-$0.0151 & $-$0.0392   & $-$0.0127\\
$L_{23}$ & $-$0.0218 & $-$0.02115  & $-$0.0350\\
\hline
\end{tabular*}
\end{table}

High resolution synchrotron x-ray diffraction has emerged as an important tool for characterizing the low
temperature structures of magnetic materials.\cite{suchomel_2012,kemei_2013,llobet_2000} However, it has
been shown that chemical inhomogeneity that is present at all times, but cannot be easily detected by high
resolution x-ray diffraction above the phase transition temperature, can lead to structural phase
coexistence below a structural phase transition temperature.\cite{llobet_2000} In this light, we have
refined the elemental occupancies of all atoms in the low temperature tetragonal and orthorhombic phases.
All elements remain stoichiometric within error. The highest vacancy concentration of 1.33\,$\%$ is
obtained for Mn$^{3+}$ cations in the orthorhombic phase. Nevertheless, this small vacancy concentration
cannot account for $>$\,50\,$\%$ orthorhombic phase content observed below $T_N$. We have performed
high-precision density measurements of the \mno\,  sample to further investigate the stoichiometry of this material.
Density measurements of the powder sample at room temperature give a density of 4.855(9) g/cm$^3$ which
compares well with the expected density of 4.86 g/cm$^{3}$ confirming that off-stoichiometric effects are
not influencing the structural phase transition of this sample.

Structural phase coexistence in \mno\, is linked to large strains at the phase transition temperature due
to the lattice mismatch between the low temperature orthorhombic phase and the high temperature
tetragonal phase. Figure. \ref{fig:LatParam} shows that the orthorhombic $a$ and $b$ lattice constants
vary by 0.3\,$\%$ from those of both the high temperature and low temperature tetragonal phases. The $c$
axis varies slightly, by 0.02\,$\%$ between the $Fddd$ and $I4_1/amd$ phases. The map of the strain
tensor shows a highly strained low temperature tetragonal phase (Fig. \ref{fig:strain}). Most of the
strain is along the crystallographic $a$ and $b$ axes where the largest lattice mismatch between the low
temperature tetragonal and orthorhombic phases is observed. The orthorhombic
$Fddd$ phase also experiences slightly higher strain compared to the high temperature tetragonal phase.
It is important to note the significantly smaller strain along the [001] direction, this finding agrees
well with the fairly invariant $c$ lattice constants of the high temperature and low temperature phases.
Evidence of strain stabilized structural phase coexistence in \mno\, is corroborated by the work of
Suzuki and Katsufuji, who have performed strain measurements on single crystals of \mno\, observing
changes in the temperature dependent strain ($\Delta L$/$L$)  at the magnetic ordering temperatures.
\cite{suzuki_2008} Suzuki and Katsufuji also show larger strain effects along the [110] plane of \mno\,
compared to the [001] direction.\cite{suzuki_2008}

Evidence of coexisting tetragonal and orthorhombic phases is in agreement with prior studies of the
hausmanite \mno. Chung $et\,al.$ reported orthorhombic instabilities in a single crystal of \mno\, after
observing subtle diffraction peak asymmetry.\cite{chung_2013} The study by Chung $et\,al.$ essentially
indicates the present of coexisting phases in \mno\, below $T_N$, with the tetragonal phase yielding
dominant peaks and the orthorhombic phase contributing to mere peak asymmetry. Considering the work of
Chung and coworkers, it is clear that large strains in a single crystal can inhibit the formation of a
large phase fraction of the low temperature orthorhombic phase. The polycrystalline sample examined here
along with the use of high-resolution synchrotron x-ray diffraction allows us to fully resolve the
diffraction reflections of the orthorhombic $Fddd$ phase and to complete a detailed study of the
structural changes taking place at low temperatures. 

\begin{figure}
\centering \includegraphics[width=3.4in]{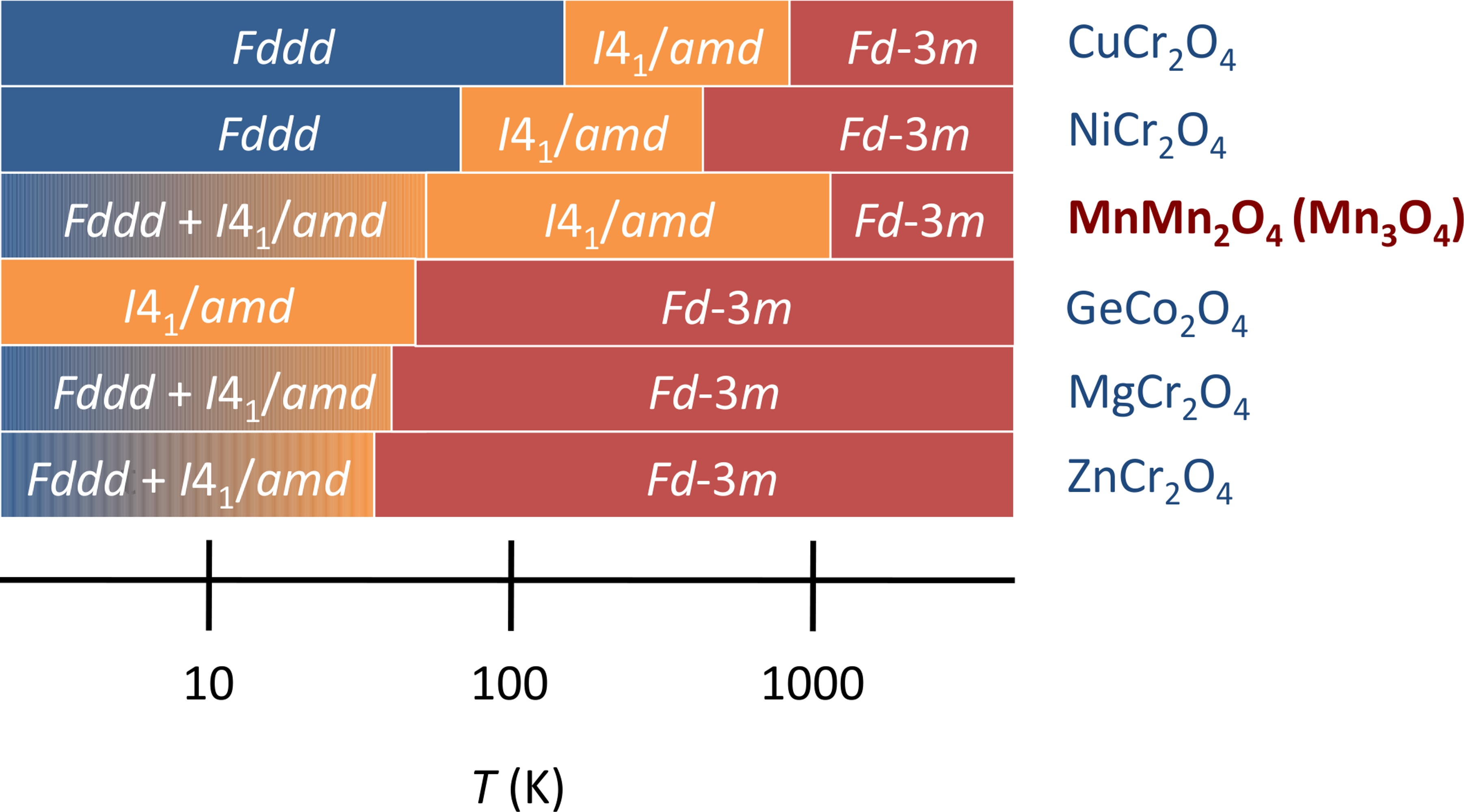}\\
\caption{(Color online) Structural distortions in some spinels due to Jahn-Teller ordering and
magnetostructural coupling effects.  While in many instances full structural transformations occur, phase
coexistence is observed for MgCr$_2$O$_4$, ZnCr$_2$O$_4$, and Mn$_3$O$_4$. 
\label{fig:spinels}}
\end{figure} 

The structural phase transformation of \mno\, bears
some of the hallmark characteristics of a martensitic phase transformation. We observe the nucleation of
an orthorhombic $Fddd$ phase within the matrix of the parent $I4_1/amd$ phase.  Strain energy inhibits
the complete structural transformation of \mno\, to the orthorhombic phase, stabilizing a mixed phase
structure. Comparable strain-mediated structural phase coexistence is reported in the manganites
Bi$_{0.2}$Ca$_{0.8}$MnO$_3$ and La$_{0.275}$Pr$_{0.35}$Ca$_{0.375}$MnO$_3$.\cite{podzorov_2001}
However, it must be noted that the perovskites are usually on the verge of being metals and the inhomogeneities
are frequently the result of very slightly differing levels of chemical doping, but this is not 
the case in \mno\, because it is stoichiometric and insulating.
Examination of hysteresis effects of the structural phase transition of \mno\, are inconclusive due to
slow heating and cooling rates applied during variable temperature x-ray measurements. A comparison of
the magnetostructural phase transformation of \mno\, to spin-driven lattice deformations of other
magnetic spinels suggests that phase coexistence likely occurs when structural deformations occur at low
temperatures ($T$\,$<$\,50\,K) (Fig. \ref{fig:spinels}). However, the spinel GeCo$_2$O$_4$ shows a full
transformation from cubic to tetragonal symmetry near 22\,K indicating that low temperatures do not
necessarily limit full structural transformations in all spinels and perhaps the particular strains
involved in the lattice deformation play a larger role in stabilizing phase coexistence. 

These results call for a re-examination of the properties of \mno\, at low temperatures. For example, how
do we understand spin ordering in the various low temperature phases? Further, it is important to resolve
the magnetic structure of \mno\, taking into account structural phase coexistence in the ferrimagnetic
phase. Since strain influences phase coexistence, it presents a new approach to controlling the
magnetostructural phase transition of \mno\, to achieve a desired low temperature structure. \\

\section{Conclusions}

High resolution synchrotron x-ray diffraction reveals the coexistence of tetragonal $I4_1/amd$ and
orthorhombic $Fddd$ below the N\'eel temperature of the magnetodielectric spinel \mno. The two low
temperature phases coexist in nearly equal fractions. A complete crystallographic description of \mno\,
in the ferrimagnetic state is presented. Polyhedral distortions in the ferrimagnetic phase of \mno\, are
described. We show that strains due to the lattice mismatch between the orthorhombic phase, which is
nucleated below 42\,K, and the high temperature tetragonal $I4_1/amd$ phase likely contribute to the
observed phase coexistence. We propose strain as a new approach to control the properties of \mno\, below
the magnetic ordering temperature. 

\subsection{Acknowledgements} 

We thank Professor Van der Ven for helpful discussions. This project was supported by the NSF through the DMR 1105301. MCK is supported by the Schlumberger
Foundation Faculty for the Future fellowship. We acknowledge the use of MRL Central Facilities which are
supported by the MRSEC Program of the NSF under Award No. DMR 1121053; a member of the NSF-funded
Materials Research Facilities Network (www.mrfn.org). Use of data from the 11-BM beamline at the Advanced
Photon Source was supported by the U.S. Department of Energy, Office of Science, Office of Basic Energy
Sciences, under Contract No. DE-AC02-06CH11357. Data were also collected on the ID31 beamline at the
European Synchrotron Radiation Facility (ESRF), Grenoble, France. We thank Andy Fitch and Caroline Curfs
for providing assistance in using beamline ID31. We thank Michael Gaultois, Quantum Design staff scientists Dr. Neil Dilley and Dr. Shi Li, and Quantum Design for high temperature susceptibility measurements. 
\bibliography{Mn3O4_Citations}
\end{document}